\documentclass[a4paper,twocolumn]{esapub} 

\usepackage{times}
\usepackage{natbib}
\usepackage{graphicx}

\makeatletter
\def\table{\@ifnextchar[{\table@i}{\table@i[\fps@table]}}
\def\table@i[#1]{\@float{table}[#1]\small}
\makeatother

\title{INTEGRAL Cross-Calibration Status}

\author{Piotr Lubi\'nski$^{1,2}$, Pierre Dubath$^{2}$, Peter Kretschmar$^{3,2}$,
Katja Pottschmidt$^{3,2}$, Roland Walter$^{2}$}
 
\affil{$^{1}$N. Copernicus Astronomical Center, Bartycka 18, PL-00716 Warszawa,
Poland}
\affil{$^{2}$INTEGRAL Science Data Centre, Chemin d'\'Ecogia 16, CH-1290 
Versoix, Switzerland}
\affil{$^{3}$Max-Planck-Institut f\"ur extraterrestrische Physik, Postfach 1312,
D-85748 Garching, Germany}

\begin{document}
 
\maketitle

\begin{abstract}
The status of the INTEGRAL cross-calibration is presented for a
standard X-ray astronomy source, the Crab Nebula, as well as for some
weaker sources. The relative flux normalization for the different
INTEGRAL instruments is discussed together with spectral shape
features.
\end{abstract}

\section{Introduction}

After one year of successful operation INTEGRAL has gathered a large
amount of scientific data. Deeper performance verification, based on
comparison of the scientific results obtained from the individual
INTEGRAL instruments and from other missions, now becomes crucial. 
This poster summarizes the current status of the cross-calibration 
with respect to spectral shape studies.

\section{Crab}

The calibration issues should be studied for a bright, steady X-ray
source with a rather simple, well established spectral model. Below we
present the comparison of INTEGRAL Crab spectra with the canonical model
of the Crab Nebula: an absorbed power law, with spectral index,
$\Gamma$ = 2.1, and with a normalization at 1 keV, $A$ = 9.7
photons cm$^{-2}$ s$^{-1}$ keV$^{-1}$ \citep{TS74}. The absorption
hydrogen column density, $N_{H}$ = $3.45\times 10^{21}$ cm$^{-2}$, was
adopted from the current XMM-Newton measurements \citep{Will01}. Spectral
modelling was done using XSPEC 11.3 \citep{Arna96}.

Fig. 1 shows the ratios between the INTEGRAL Crab data and the
canonical model. Spectra from revolution 102 (15-17 August 2003) were
extracted with the OSA 3.0 software. For all instruments the absolute
normalization is still not perfect. The result for JEM-X 2 is not
satisfying below 7 keV. This is due to an incorrect flux
reconstruction for lower energies and makes the spectra appear more
absorbed than they actually are. The ISGRI case is the worst, with the
spectrum being distorted by 'snake-like' features below $\approx$100
keV and with declining flux above 200 keV. The SPI spectrum, except
for normalization, agrees well with the model, being quite smooth in
the 30-200 keV range.
	
\begin{figure}
\centering
\includegraphics[width=\linewidth]{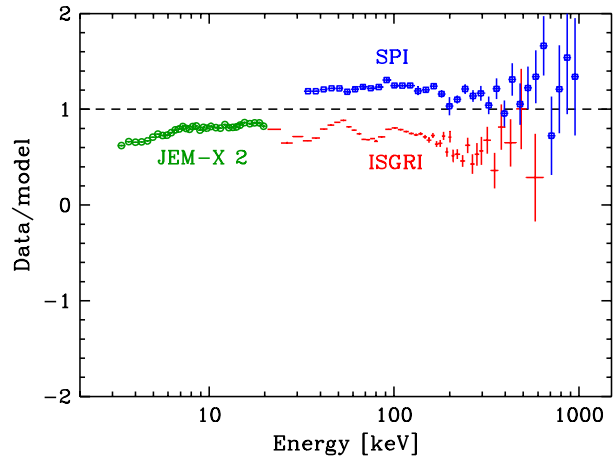}
\caption{INTEGRAL data for Crab from revolution 102 compared to the canonical 
model.\label{fig:crabratio}}
\end{figure}
		
Fixing the canonical model parameters we have fitted the relative
normalization parameters for the INTEGRAL instruments. (JEM-X 2 spectra
are always fitted in the 7-21 keV range in the following, while the energy 
range used in fitting for ISGRI and SPI is that shown in figures.) The fit is
bad and releasing the power law index does not improve it. For
individual instruments we obtain almost acceptable fits for JEM-X 2
and SPI, but with a model which is a little too flat in the JEM-X 2
case. The ISGRI model gives a slope equal to the canonical value but
the fit is poor. All these results are listed in Table 1. During
fitting we have applied only statistical errors, except for ISGRI, for
which we have tested an alternative fit, after adding systematical
errors of 10\% to the data. This leads to an acceptable $\chi^{2}$ and
a quite good absolute normalization but with a larger slope.

It should be mentioned that the ISGRI spectra shown here were
extracted for science windows with the source being on axis. Selecting
data with the object in the fully coded field of view gives a worse,
lower normalization, about 50\% of that obtained for on-axis spectra.

In Fig. 2 we present the comparison between the Crab spectra extracted
with the OSA 3.0 and OSA 2.0 packages. The data for SPI are not shown,
since for SPI there was no change in the software affecting the
calibration in these two OSA releases. Both JEM-X 2 and ISGRI data
from OSA 2.0 are farther from the proper absolute
normalization. However, with OSA 3.0 the 'absorption' feature
mentioned above arises for JEM-X 2. The scattering of the ISGRI flux
is suppressed for OSA 3.0 in comparison to OSA 2.0 but the overall
spectral shape remains bad. Please note the good agreement between
the absolute JEM-X 2 and ISGRI fluxes at 20 keV with OSA 3.0
	
\begin{figure}[h!]
\centering
\includegraphics[width=\linewidth]{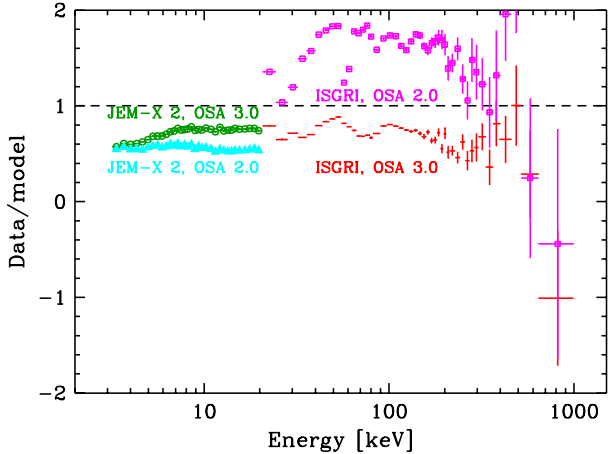}
\caption{INTEGRAL Crab data from revolution 102 divided by the
canonical model, comparison between spectral extraction done with OSA
2.0 and OSA 3.0.
\label{fig:single}}
\end{figure}

\tabcolsep 2pt
\begin{table}[h!]
\begin{center}
\caption{Results of spectral fitting obtained for Crab spectra from
INTEGRAL revolution 102. C values are the relative normalization
factors, A and $\Gamma$ are parameters of the power law model. The second 
fit for ISGRI was done after adding systematic errors of 10\% to the
data. Fixed parameters are denoted with f.}
\vspace{1em}
\renewcommand{\arraystretch}{1.2}
\begin{tabular}[h]{lcccccc}
\hline
Instrument & $C_{JEM-X 2}$ & $C_{ISGRI}$ & $C_{SPI} $ & A & $\Gamma$ & 
$\chi ^{2} /$NDF \\
\hline
INTEGRAL     & 0.82 & 0.75 & 1.21 & 9.7f & 2.1f & 9320/98 \\
INTEGRAL     & 0.80 & 0.73 & 1.18 & 9.7f & 2.09 & 9302/97 \\
JEM-X 2	     & 1.0  & -    & -    & 6.95 & 2.05 & 41.8/19 \\
ISGRI        & -    & 1.0  & -    & 7.22 & 2.10 & 9146/43 \\
ISGRI(10\%) & -    & 1.0  & -	  & 10.4 & 2.19	& 48/43 \\
SPI          & -    & -    & 1.0  & 10.8 & 2.08 & 82.5/33 \\
\hline \\
\end{tabular}
\label{tab:crabtable}
\end{center}
\end{table}

\section{Weaker sources}

Since INTEGRAL spectra are background dominated, the spectral
extraction for weak sources becomes more dependent on the background
model. Therefore, cross-calibration tests made for a strong source,
such as the Crab, obviously will not exhaust all the issues of the
INTEGRAL calibration. Below we present a review of a simple study of
the INTEGRAL spectral shapes obtained for sources with relatively
medium (Cyg X-1) or weak (3C 273, NGC 4151) fluxes.

As these sources exhibit some spectral variability, there is no canonical
model for them and the tests are more difficult than in the Crab case. 
For comparison's sake we adopted as a reference the data collected by RXTE 
simultaneously with the INTEGRAL observations. We used
the XSPEC models {\tt \small constant*phabs
(diskbb+comptt+reflect(comptt)+gauss)} for Cyg~X-1, {\tt \small
constant*wabs *powerlaw} for 3C~273 and {\tt \small
constant*wabs*absnd*(cutoffpl+zgauss)} for NGC 4151 ({\tt \small absnd} is
a local model for partial covering absorption). The relative
normalization was fixed at 1.0 for HEXTE, we adopted $N_H$ values from
the HEASARC nH column density tool, and we used solar
abundances. Figs. 3-5 present the ratios between the INTEGRAL data and
the best fit RXTE model for Cyg X-1, NGC 4151, and 3C 273,
respectively.
	
\subsection{Cyg X-1}
	
Cyg X-1 was observed by INTEGRAL during revolution 11 (16-18 November 2002) 
and by RXTE on 16 November 2002. The ratios between the INTEGRAL data and 
the RXTE model obtained for this object are similar to those shown in Figure 
1 for the Crab. This is
not surprising since the Cyg X-1 2-20 keV flux is equal to 1.1 photons
cm$^{-2}$ s$^{-1}$, i.e., comparable with the Crab flux of 2.8 photons
cm$^{-2}$ s$^{-1}$. Nevertheless, the SPI flux in the 150-300 keV
range is lower than the corresponding values for Crab. For ISGRI the
data/model ratio resembles that of the Crab, with deformations
below 100 keV and with declining flux for higher energies.
		
\begin{figure}[h!]
\centering
\includegraphics[width=\linewidth]{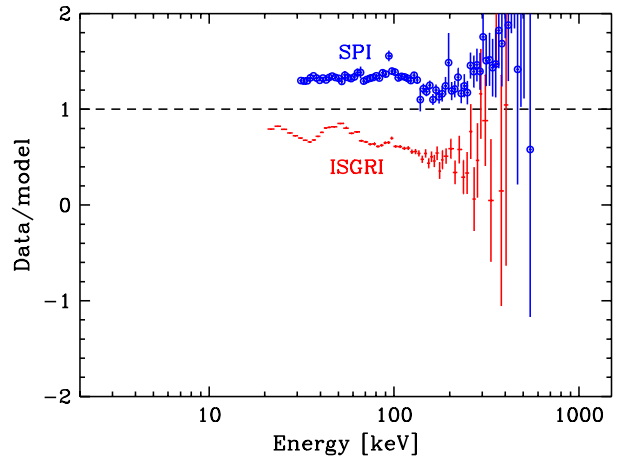}
\caption{INTEGRAL data for Cyg X-1 from revolution 11 compared to the
RXTE best fit model. A JEM-X 2 spectrum was not extracted due to its
non-standard operation mode for this
revolution. \label{fig:cygx1ratio}}
\end{figure}
		
\subsection{NGC 4151}
			
INTEGRAL observation of NGC 4151 was performed during revolutions 74-76 
(23-29 May 2003). Corresponding RXTE campain was shorter: useful data for 
PCA are from the period 24-29 May 2003 and for HEXTE only from 27-29 May 2003. 
The 2-20 keV flux fitted to the PCA data of NGC 4151 is equal to 0.045
photons cm$^{-2}$ s$^{-1}$. For an object about 100 times fainter than
the Crab the overall shapes of the JEM-X 2 and ISGRI data-to-model
ratios are still similar to the Crab results, however, with a different 
normalization, very close to 1. This apparent agreement of the INTEGRAL
instruments absolute flux with the HEXTE results can be explained by the 
fact that HEXTE data were collected during two short periods of INTEGRAL 
observation, when the source flux was, on average, about 20\% lower than 
the flux measured during all 7 days.
		
\begin{figure}[h!]
\centering
\includegraphics[width=\linewidth]{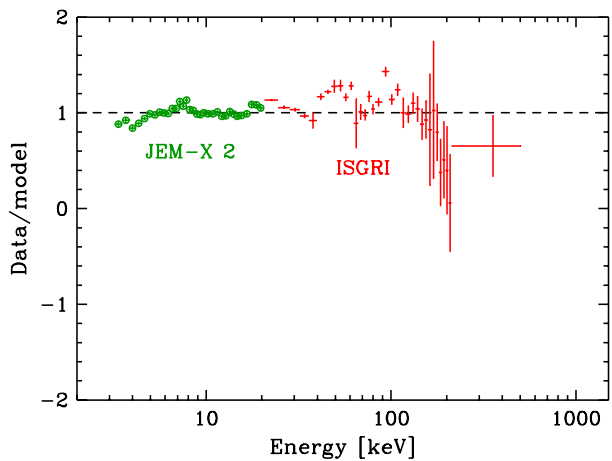}
\caption{INTEGRAL data for NGC 4151 from revolutions 74-76 compared to
the best fit RXTE model. Due to the staring observation strategy a
spectrum for SPI was not extracted. \label{fig:ngc4151ratio}}
\end{figure}
		
\subsection{3C 273}

3C 273 was observed simultaneously by INTEGRAL (revolution 82) and RXTE 
on 16 June 2003. This is the weakest source in our sample, with a 2-20 keV 
flux equal to 0.027 photons cm$^{-2}$ s$^{-1}$, and due to this fact its
results are less conclusive. The relative normalizations obtained for
the INTEGRAL instruments are clearly lower than the corresponding
values derived for stronger sources. This can possibly be explained by
the background model being worse. Again, the JEM-X 2 and ISGRI spectra
descend too much on their lower and higher energy boundaries,
respectively. The decrease of the ISGRI flux with increasing energy
can be explained by the inefficiency of the standard extracting
method, working on individual science windows. As can be seen from
Figs. 1, 3-5, this effect correlates with the decreasing source
strength. The alternative ISGRI spectrum (O) extracted in a
non-standard way, from the mosaic image, looks better than the
standard (S), being almost undeformed below 100 keV and without
steepening above 100 keV. However, its absolute normalization is
somewhat worse than in the ISGRI (S) case. The SPI data are uncertain
and probably exhibit some background contamination because of the
presence of strong instrumental lines in this range.

\begin{figure}
\centering
\includegraphics[width=\linewidth]{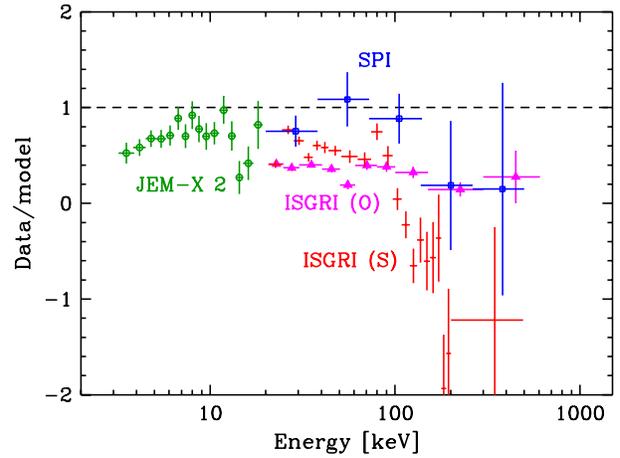}
\caption{INTEGRAL data for 3C 273 from revolution 82 compared to the
RXTE best fit model. For ISGRI an alternative approach to the spectral
extraction is presented, where the overall observation spectrum (O) is
prepared from the mosaic image, instead of summing the spectra
obtained for individual science windows (S).
\label{fig:3c273ratio}}
\end{figure}

For the details of the spectral modelling for these three sources we
refer the reader to the posters devoted to them. For this review we
have only fitted the relative normalization for INTEGRAL instruments,
with the rest of model parameters fixed at the RTXE values. The
results are gathered in Table 2.  As stated before, the absolute
normalization for weaker objects is still uncertain. (We do not quote
$\chi^{2}$ values for these results since they are not relevant for
such a limited fit.)

\tabcolsep 3pt
\begin{table}[h!]
\begin{center}
\caption{Relative normalization factors fitted for INTEGRAL spectra of Cyg X-1, 
3C 273 and NGC 4151, with the other parameters fixed at the best fit values 
obtained for RXTE data. For ISGRI (S) we show also the value obtained for 
spectrum below 100 keV.}\vspace{1em}
\renewcommand{\arraystretch}{1.2}
\begin{tabular}[h]{lcccc}
\hline
Object & $C_{JEM-X 2}$ & $C_{ISGRI}$ (S) & $C_{ISGRI}$ (O) & $C_{SPI}$ \\
\hline
Cyg X-1   & - & 0.74 & - & 1.32 \\
3C 273    & 0.72 & 0.52/0.56 & 0.37 & 0.82 \\
NGC 4151 & 1.01 & 1.11 & - & - \\
\hline \\
\end{tabular}
\label{tab:weaktable}
\end{center}
\end{table}

\section{Conclusions}

INTEGRAL spectra extracted currently should be treated with caution. At the 
present time none of the INTEGRAL instruments can be used to measure the 
absolute flux. The major problem of JEM-X 2 with respect to the
spectral shape is the artificial absorption excess observed below 7
keV. The response matrices for ISGRI are still preliminary and the
spectra are too deformed to be used for detailed spectral
modeling. Only SPI does not exhibit any serious problems in its
spectral shape calibration. The calibration of the INTEGRAL
instruments is in progress, the new major release 4.0 of OSA should solve
many of the issues discussed here.

\end{document}